\documentclass[%
 reprint,
 amsmath,amssymb,
 aps,
 pra,
]{revtex4-2}

\usepackage[nottoc]{tocbibind}
\usepackage[english]{babel}
\usepackage[utf8]{inputenc}
\usepackage{blindtext}
\usepackage[T1]{fontenc}
\usepackage{cool}
\usepackage{titlesec}
\usepackage{amsmath}
\usepackage{physics}
\usepackage{mathtools}
\usepackage{graphicx}
\usepackage[labelformat=simple]{caption}
\usepackage{float}
\usepackage{color}
\usepackage{yfonts}
\usepackage{dcolumn}
\usepackage{bm}
\usepackage{caption}
\newcommand{\zbar}{\raisebox{-0.01ex}{-}\kern-0.4em z}
\usepackage{gensymb}

\begin{document}

% \title{Controlling Wave Transmission Through Ultra Thin Networks}

% \title{Which Waveguide Network  \\ An Exact Forward Construction}

% \title{Exact Forward Construction of Ideal Wave Transmission Through Waveguide Networks}

\title{Which Waveguide Network Realizes a Prescribed Transmission Profile? \\ An Exact Forward Construction}

% \author{T. M. Lawrie$^{1,2}$\footnote{Corresponding author: \href{mailto:tristan.lawrie@nottingham.ac.uk}{tristan.lawrie@nottingham.ac.uk}}, G. Tanner$^1$, G. J. Chaplain$^2$}

\author{T. M. Lawrie$^{1*}$}

\affiliation{$^{1}$Centre for Metamaterial Research and Innovation, Department of Physics and Astronomy, University of Exeter, United Kingdom}

\email{ \href{mailto:tristan.lawrie@nottingham.ac.uk}{t.lawrie@exeter.ac.uk}}

\date{\today}% It is always \today, today,

% \title{A Novel Angular Filter Device - Via Quantum Graph Theory}
% \author{Tristan\ Lawrie$^1$, Gregor\ Tanner$^1$\\
% {\normalsize School of Mathematical Sciences, University of Nottingham$^{1}$;}
% \date{\today}}

% \begin{document}

\begin{abstract}

We introduce an analytically invertible framework for wavefront construction based on the scattering properties of periodic waveguide networks governed by a gauge-shifted Helmholtz operator. By determining the exact transmission coefficients of the network, we express the lattice reactance as a Fourier expansion whose coefficients are analytically mapped onto the underlying graph architecture, allowing the required bond connections, refractive indices, lengths, and gauge phases to be determined directly from a prescribed target transmission coefficient. In contrast to conventional inverse-design approaches, the present formulation provides a closed-form route from desired wave transmission profiles to physically realisable structures. The framework extends naturally from one-dimensional angular filtering to two-dimensional image synthesis, where arbitrary transmitted intensity patterns are reconstructed through exact spectral control of the network scattering response.

% Source $x = -L$

% Filter $x = 0$

% Image $x = L$

% $X$

% $Y$

% $Z$

% $n$

% $m$

% $\ell_{0,2}$

% $\ell_{1,0}$

% Ideal Transmission

% Constructed Transmission 

% Pruned Graph Transmission

% $\epsilon$

% $\ell$

% $k_y\ell$

% $|t_p|^2$

% Connection Number $p$

% Bond Length $\ell_p$

% Refractive index $n_p$

% Magnetic Phase $\phi_p$

% $|T|^2$

% $k_y$

% $1/n_p$

% $\ell_p$

% $\phi_p$

% $p$

% $X = n\ell$

% $Y = m\ell$

% $|\psi_{n,m}|^2$

% Ideal

% Constructed

% Pruned

% Reduced Valency 

% Kept

% Pruned

% Tol

% (a)

% (b)

% (c)

% (d)

% (e)

% Ideal Transmission $|T(k_y)|^2$

% Constructed from 16 Modes

% Constructed from 9 modes 

% (a) (b) (c) (d) (e) (f) (g)

% $p$

% $q$

% $k_y$

% $k_z$

% $1/n_{p,q}$

% Pruned $1/n_{p,q}$

% $\ell_{p,q}$

% Pruned $\ell_{p,q}$

% $\phi_{p,q}$

% Pruned $\phi_{p,q}$

% $|T|^2$ Full Construction

% $|T|^2$ Pruned Construction

\end{abstract}

% \begin{abstract}
% In the present work, we study the transmission properties of a 1 and 2 dimensional angular filter. We show that given some ideal transmission image, the exact filter properties can be trivially determined via an inverse Fourier Transform. 
% \end{abstract}

\maketitle

\begin{figure}[t]
    \centering
    \includegraphics[width=0.9\linewidth]{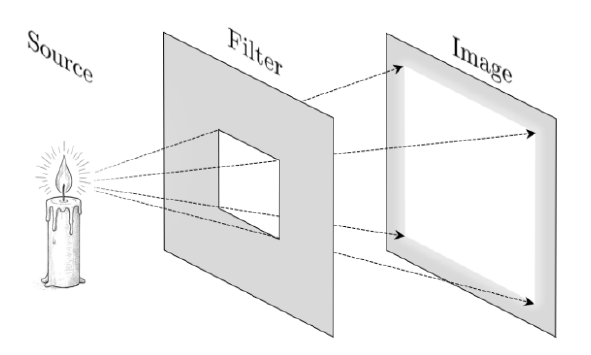}
    \caption{Illustrates a traditional light filter, where the cut out section produces an image of the same shape. The conclusion of this paper shows an ideal reconfigurable version of the same effect for an arbitrary scaler wave field. We consider the wavelength of the emitted light to be much smaller than the characteristic dimensions of the filter, so diffraction effects are negligible.}
    \label{fig: Idea}
\end{figure}

\begin{figure*}[t!]
    \centering
    \includegraphics[width=0.97\linewidth]{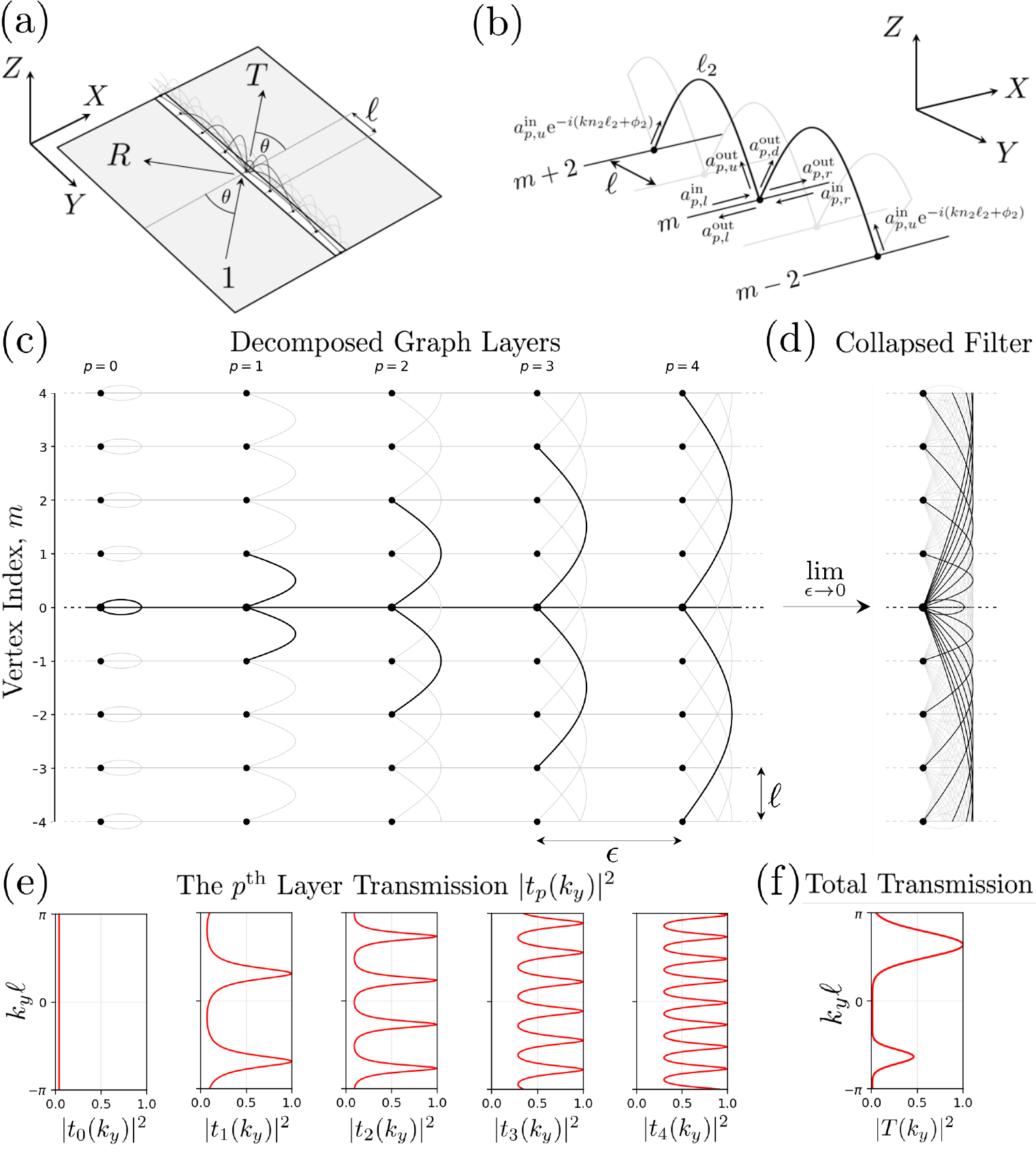}
    \caption{(a) A filter formed from many internal graph connections couples two semi-infinite scattering regions, shown in grey, in which waves propagate freely. An incident wave of unit amplitude and angle \(\theta\) scatters from the filter into reflected and transmitted components with amplitudes \(R\) and \(T\), respectively. (b) Example of a single connection order, here \(p=2\), in which vertex \(m\) is coupled to vertices \(m\pm2\) by graph bonds of length \(\ell_2\). The local wave amplitudes incident on and outgoing from vertex \(m\) are indicated. (c) The full filter may be represented as a stack of Fourier graph layers, with each layer containing one connection order \(p=0,1,2,\ldots\), separated by a small spacing \(\epsilon\). (d) The physical collapsed filter is obtained by taking the zero-spacing limit \(\epsilon\to0\), yielding a single composite graph filter containing all connection orders. (e) Single-layer transmission intensities \(|t_p(k_y)|^2\) for the first few Fourier graph layers. (f) Resulting transmission intensity \(|T(k_y)|^2\) of the collapsed composite filter.}
    \label{fig:Graph Setup}
\end{figure*}

\section{Introduction}

\begin{table*}[t]
\centering
\caption{The gauge shifted one dimensional time independent wave equation in different wave systems.}
\label{tab:wave_mapping}
\renewcommand{\arraystretch}{1.15}
\begin{tabular}{c|c|c|c}
\hline
\textbf{Physical System} &
\textbf{Wave Field} &
\textbf{Gauge / Flow Term} &
\textbf{Physical Interpretation}
\\
\hline

Quantum Mechanics
&
\(
\psi \rightarrow \psi_q
\)
&
\(
A \rightarrow \mathbf{A}_{\mathrm{mag}}
\)
&
\shortstack[c]{Magnetic vector potential\\
producing Aharonov--Bohm phase accumulation}
\\

\hline

Acoustics with Flow
&
\(
\psi \rightarrow \phi
\)
&
\(
A \rightarrow U
\)
&
\shortstack[c]{Background flow velocity\\
producing convective phase bias}
\\

\hline

Microwave / Coaxial Networks
&
\(
\psi \rightarrow u
\)
&
\(
A \rightarrow \beta_{\mathrm{nr}}
\)
&
\shortstack[c]{Non-reciprocal propagation phase\\
generated through modulation or biased components}
\\

\hline

Photonics / Optical Networks
&
\(
\psi \rightarrow E
\)
&
\(
A \rightarrow \Delta \beta
\)
&
\shortstack[c]{Synthetic gauge phase generated through\\
modulation, moving media, or refractive-index bias}
\\

\hline

Mechanical Metamaterials
&
\(
\psi \rightarrow \eta
\)
&
\(
A \rightarrow V_{\mathrm{bias}}
\)
&
\shortstack[c]{Directed elastic-wave transport induced\\
through rotation, motion, or active bias}
\\

\hline
\end{tabular}
\end{table*}

The ability to prescribe the propagation of waves through a material lies at the heart of holography, image formation, beam shaping, beam steering and wave-based information processing. Since Gabor's original formulation of holography \cite{gabor1948} and the development of iterative phase-retrieval methods such as the Gerchberg--Saxton algorithm \cite{gerchberg1972}, a central problem has been to determine the physical structure required to generate a desired image or wavefront. Metamaterials and metasurfaces have become powerful platforms for this purpose, enabling compact control over phase, amplitude, polarisation and dispersion across optical, microwave, terahertz and acoustic systems \cite{yu2011,kildishev2013,padilla2022}. Recent developments in nonlocal and diffractive metasurfaces further show that spatially extended resonant modes can be used to generate highly selective wavefront responses, including Fano-resonant and bound-state-in-the-continuum-based devices \cite{OvervigYuAlu2021}.

Despite this progress, image-forming and wavefront-synthesis devices are still most commonly designed through inverse procedures. Computer-generated holography, meta-optics and metasurface imaging typically prescribe the desired output field and subsequently determine the physical architecture through phase retrieval, non-convex optimisation, adjoint methods, topology optimisation, surrogate modelling or machine-learning-assisted design \cite{sui2024nonconvex,wang2024phaserecovery,li2022largescale,yin2024multidimensional,choi2024largearea,zeng2025imagingmetasurface,zhang2025multichannel,park2025thirtysixchannel,yang2025aimetasurface,yu2025deeplearningcgh}. These methods have enabled remarkable progress in large-area meta-optics, multiplexed holography and computational imaging, but the underlying philosophy remains the same: the desired wave response is specified first, while the corresponding material architecture is obtained through optimisation, training or repeated solution of a forward scattering problem.

A closely related development is analogue wave computing, where mathematical operations are performed directly on propagating fields rather than after digital acquisition. Metamaterials and nanophotonic structures have been shown to implement operations such as spatial differentiation, integration, convolution, Fourier processing and image filtering \cite{Silva14,Alu21,Wesemann21,Guo18}. Diffractive optical neural networks extend this idea further by embedding trained computation directly into passive wave propagation \cite{lin2018alloptical}. These works demonstrate that engineered media can act not only as wavefront shapers, but as physical operators acting on incident fields. In this context, the design problem may be viewed as the construction of a desired transmission operator.

A common route to modelling metamaterials is through spring--mass or beam lattices, where the wave properties of the medium are understood through its band structure. Within this setting, nonlocal or beyond-nearest-neighbour couplings play a central role \cite{Brillouin1960}. Such interactions have been shown to generate roton-like dispersion relations, multiple propagating modes at a single frequency, anti-parallel phase and group velocities, surface modes and higher-order degeneracies in acoustic, elastic and mechanical metamaterials \cite{chen2021roton,iglesias2021experimental,wang2022nonlocal,chen2023observation,chaplain2023reconfigurable,moore2023acoustic,fleury2021non,edge2025discrete}. These results establish nonlocality as a powerful route to dispersion control beyond that available in locally coupled lattices.

A particularly important step was taken by Kazemi \emph{et al.}, who showed that prescribed one-dimensional dispersion relations can be constructed directly using nonlocal phononic crystals \cite{kazemi2023drawing,kazemi2023non}. In this framework, the target band structure is written as a Fourier series and the Fourier coefficients determine the required beyond-nearest-neighbour spring constants. This exact-synthesis philosophy has since been extended to two-dimensional dispersion surfaces \cite{Wang2024}, complex-valued dispersion relations \cite{edge2025engineering}, and related beyond-nearest-neighbour beam-lattice architectures \cite{edge2025discrete}. These works show that, within closed spring--mass or beam-lattice models, band engineering may be reformulated as an explicit synthesis problem rather than a numerical inverse-design problem.

The present work transfers this philosophy from closed lattice models to open waveguide networks. The platform considered here is a network of thin waveguides, often referred to in the mathematical physics literature as a quantum graph. Quantum graphs provide a compact scattering framework for wave propagation on networks and have been widely studied in relation to spectral theory, quantum chaos, open transport and wave communication \cite{kottos1999periodic,gnutzmann2006quantum,berkolaiko2013introduction,barra2001transport,hein2009wave,lawrie2023closed}. They also arise naturally as the thin-waveguide limit of graph-like domains \cite{kuchment2001convergence,rubinstein2001variational, exner2005convergence}, making them a physically meaningful model for acoustic, microwave and mechanical waveguide networks.

Quantum graph theory has recently been used as a design framework for metamaterial interfaces and waveguide-network devices. Lawrie \emph{et al.} showed that graph-based metamaterials can be used to engineer interface scattering coefficients \cite{lawrie2023engineering}, negative refraction of acoustic waveguide modes \cite{lawrie2024application}, and non-diffracting resonant angular filtering \cite{lawrie2025nondiffracting}. Subsequent work introduced magnetic phases through the magnetic Schrödinger operator, making the angular pass direction continuously tunable \cite{lawrie2025flux}. Related discrete angular filtering has also been experimentally realised using microwave coaxial networks \cite{yv76-8rw6}. These results show that graph architectures provide a physically realisable counterpart to abstract nonlocal lattice models, with bonds corresponding to actual waveguides and vertices corresponding to experimentally accessible junctions.

Here we show that the same beyond-nearest-neighbour architecture does more than engineer the closed-system band diagram of a lattice. When the graph is coupled to a scattering environment, the internal graph architecture determines the transmission coefficient of waves incident on the material. In particular, the transmission coefficient of the waveguide network naturally takes the form of a Fourier series, where each harmonic corresponds to a graph connection and each coefficient is fixed by the bond refractive index, bond length and magnetic phase.

This establishes an exact forward construction for ideal transmission. Rather than asking which transmission profile is produced by a chosen network, we prescribe the desired transmission profile and determine the waveguide architecture directly. In this sense, the work shifts the perspective from closed-system band engineering to open-system transmission engineering, providing a physically realisable waveguide-network route to holography, image formation, wavefront synthesis, analogue wave computing and beam steering without iterative inverse design.

The paper is organized as follows. Section~\ref{sec: A Standard Representation} expresses ideal transmission through a thin interface in terms of an effective reactance. Sections~\ref{sec: Modelling the Filter via Quantum Graph Theory} and~\ref{sec: Scattering from a filter with an infinite number of internal connections} show that a waveguide network realizes this reactance through a Fourier series generated by its internal architecture. Section~\ref{sec: Fourier Representation} gives the resulting forward construction: the prescribed transmission profile fixes the graph refractive indices, bond lengths and magnetic phases directly. Section~\ref{sec: Extension to 2D} extends the construction to two-dimensional filters and image construction, before the main conclusions are summarized in Section~\ref{sec: Conclusion}.

\section{Transmission Through an Ideal Interface}\label{sec: A Standard Representation}

A standard problem in wave physics is to determine the boundary conditions required to produce some prescribed transmission coefficient. The issue arises that, yes indeed one can state the boundary conditions, but typically one cannot build a physical device that would possess such conditions. In this work, we show that indeed any prescribed boundary conditions can be achieved by coupling the two half spaces via some network of thin waveguides. 

To begin, we determine the required boundary conditions of an interface between two semi-infinite half spaces in terms of some ideal transmission coefficient. Suppose that two semi-infinite half-spaces
\begin{equation}
\begin{split}
\Omega_- &= {(x,y)\in\mathbb{R}^2 : x<0},\\\
\Omega_+ &= {(x,y)\in\mathbb{R}^2 : x>0},
\end{split}
\end{equation}
are separated by an interface, or filter, located at \(x=0\). In each region the wave field satisfies the Helmholtz equation
\begin{equation}
(\nabla^2+k^2)\phi(x,y)=0.
\end{equation}
We assume that the interface possesses an effective response characterised by a parameter \(\alpha(k_y)\), which may depend upon the transverse wave number \(k_y\). 
Such a boundary must satisfy the following two conditions,
\begin{enumerate}
    \item the wave functions are continuous at the interface
    \begin{equation}\label{eq:continuity}
        \psi (0^{+}, y) = \psi (0^{-}, y);
    \end{equation}
    \item the normal derivative satisfies the Robin jump condition,
    \begin{equation}
\frac{\partial \phi}{\partial x}(0^+,y)-
\frac{\partial \phi}{\partial x}(0^-,y)
=-k^2\alpha(k_y)\phi(0,y),
\end{equation}
\end{enumerate}
The parameter \(\alpha(k_y)\) may then be chosen to characterizes the effective response of the interface. The field in \(\Omega_-\) is expressed as a superposition of an incident wave with amplitude $1$, and a reflected plane wave with amplitude $R$,
\begin{equation}
\phi_-(x,y)
=e^{i(k_xx+k_yy)}
+
R(k_y)e^{i(-k_xx+k_yy)}.
\end{equation}
The field in $\Omega_+$ is simply the transmitted wave with amplitude $T$,
\begin{equation}
\phi_+(x,y)
=T(k_y)e^{i(k_xx+k_yy)}.
\end{equation}
Here $k_x=\sqrt{k^2-k_y^2}$.
Continuity of the field immediately yields
\begin{equation}
T(k_y)=1+R(k_y).
\end{equation}
Substituting the scattering ansatz into the Robin jump condition allows us to express the coefficient $\alpha$ in terms of the transmission coefficient,
\begin{equation}
F(k_y)
:=
i\left(
\frac{1}{T(k_y)}-1
\right)
=\frac{k^2\alpha(k_y)}
{2k_x}.
\end{equation}
Here the function $F$ is defined as the boundary reactance which relates to the transmission coefficient as, 
\begin{equation}\label{Transmission in terms of Reactance}
T(k_y) = 
\frac{1}{1-iF(k_y)}.
\end{equation}
If \(F(k_y)\) is real-valued, the interface is passive (lossless). In this case, $|R(k_y)|^2+|T(k_y)|^2=1$.
For a passive interface, \(F(k_y)\) is a real-valued function that can be written as a Fourier series. This structure completely characterises the angular transmission response which allows us to formulate the problem of constructing a physical structure whose effective reactance is equal to a prescribed function \(F(k_y)\).

The remaining question is therefore whether a physical system exists whose scattering properties realise an arbitrary reactance function. In the following sections we show that periodic quantum graph networks provide exactly such a construction. By exploiting the analytic scattering properties of the graph, it becomes possible to synthesise a prescribed reactance function directly through the graph topology, bond lengths, refractive indices and magnetic phases. The graph therefore acts as a physically realisable impedance surface whose effective boundary condition reproduces a desired transmission coefficient.

\section{Modelling the Filter via Quantum Graph Theory}\label{sec: Modelling the Filter via Quantum Graph Theory}

The aim of this section is to construct the refection $R$ and transmission coefficient $T$ of a network of thin waveguides as illustrated in Fig.\ref{fig:Graph Setup} (a). The 1D filter possesses an infinite number of connections, illuatrated in (d) with the decomposition into discrete layers in (c). Each layer of the filter has index $p$, with the resulting transmission coefficient $t_p$ illustrated below each layer in (e). An example of connection $p = 2$ is given in (b). We consider each of the waveguides to be vanishingly thin, such that it only supports the fundamental mode. In this limit the network describes a metric graph endowed with a $1D$ wave equation \cite{lawriethesis, lawrie2022quantum}. 

Let us determine the transmission coefficient $t_p$ of a given layer $p$. Consider a metric graph $\Gamma(\mathcal{V},\mathcal{E},L)$ embedded in $\mathbb{R}^3$ constructed from an infinite set of vertices $\mathcal{V}$ placed periodically along the $Y$-axis with spacing $\ell$. Each vertex is referenced by a discrete index $m\in\mathbb{Z}$, coupled by an infinite set of bidirectional edges $\mathcal{E}$ each with metric length $L = \{l_e \in \mathbb{R}^+ \mid e \in \mathcal{E} \}$. The set of edges with finite length will be called bonds $\mathcal{B}$, while the set of edges with semi-infinite length will be called leads $\mathcal{L}$, with the complete set of edges given by the union $\mathcal{E}=\mathcal{L} \cup \mathcal{B}$.

Each vertex $m$ is coupled up ($u$) and down ($d$) to vertices $m\pm p$ for $p \in \mathbf{Z}^{+}$ by bonds of length $\ell_u = \ell_d :=\ell_p$. The graph is made open by coupling leads along the $x$-axis to the left ($l$) and right ($r$) of the vertex, which act as channels to the scattering environment - see Fig. \ref{fig:Graph Setup} (b). The set of edges coupled to a given vertex $m$ form a subset of $\mathcal{E}$ which we define as the star of the vertex $\mathcal{S}_m= \{l_m,r_m,d_m,u_m\}$. For each edge $e\in\mathcal{S}_m$, we introduce an edge coordinate $z_{m,e}$ with origin chosen to be at vertex $m$, which spans the domain $z_{m,l} \in [0,\infty)$, $z_{m,r} \in [0,\infty)$ and $z_{m,d} \in [0,\ell_p]$, $z_{m,u} \in [0,\ell_p]$. Note the use of $z$ as an edge coordinate rather than the traditional Euclidean coordinates $X,Y$ and $Z$.

The metric graph is turned into a quantum graph by imposing a wave equation on each edge, as well as enforcing a choice of boundary conditions on each vertex. Here we consider the one-dimensional magnetic Schrödinger equation,
\begin{equation}\label{eq:schrodinger}
    \left(
    - i \frac{\partial}{\partial z_{m,e}}
    + A_{m,e}
    \right)^2
    \psi_{m,e}(z_{m,e})
    =
    k_e^2
    \psi_{m,e}(z_{m,e}),
\end{equation}
where the real constant \(A_{m,e}\) represents a magnetic potential along edge \(e\) connected to vertex \(m\). We assume that the leads carry no magnetic contribution, such that \(A_{m,l}=A_{m,r}=0\), while the oriented bonds carry opposite magnetic potentials,
\begin{equation}
A_{m,d}=-A_{m,u}.
\end{equation}

The chosen operator may be viewed as a gauge-shifted extension of the standard Helmholtz equation and therefore appears across a wide range of wave systems beyond quantum mechanics -  Table~\ref{tab:wave_mapping} for the correspondence between these physical systems. In each case, the governing equation may be interpreted as a gauge-shifted Helmholtz operator.

A given edge wave number $k_e$ is expressed in terms of the free-space wave number $k$ via a bond refractive index $n_e$ such that $k_e = n_e k$. As the leads act as scattering channels to free space we set
\begin{equation}
n_l = n_r = 1,
\end{equation}
and we state that the bond refractive indices are the same when traveling up and down the filter, 
\begin{equation}
n_d = n_u:=n_p.
\end{equation}

The general solution of equation~(\ref{eq:schrodinger}) is a superposition of counter-propagating plane waves, multiplied by a Bloch phase and a magnetic phase,
\begin{equation}\label{eq:schrodingersolution}
    \psi_{m,e} = \text{e}^{i(k_ym\ell - A_{m,e}z_{m,e})} \left( a^{\text{out}}_{e} \text{e}^{i k_e z_{m,e}} + a^{\text{in}}_e \text{e}^{- i k_e z_{m,e}} \right).
\end{equation}
Here, $a^{\text{out/in}}_e$ are the complex wave amplitudes heading out of or into a given vertex on edge $e$ and is illustrated in Fig. \ref{fig:Graph Setup} (b). The Bloch phase $\text{e}^{ik_y m \ell}$ arises as a consequence of the lattice periodicity with Bloch wave number $k_y$.

We choose the vertex boundary conditions to be Kirchhoff-Neumann, which under gauge transformation,
\begin{equation}\label{eq: Transform}
    \psi_{m,e} (z_{m,e}) \rightarrow \Phi_{m,e}(z_{m,e})
    =
    \text{e}^{i \int_{0}^{z_{m,e}} A_{m,e} d z_{m,e}}
    \psi_{m,e}(z_{m,e}),
\end{equation}
take the recognizable form, that being:
\begin{enumerate}
    \item Continuity of the wave function across the vertex star (waveguide junction)
    \begin{equation}\label{eq:gaugetransformedcontinuity}
    \Phi_{m,e} (0) = \Phi_{m,e'} (0),
\end{equation}
    \item The total wavefunction gradient sums to zero at the vertex star, \begin{equation}\label{eq:gaugetransformedkirchoff}
    \sum_{e \in \mathcal{S}_m} \frac{\partial \Phi_{m,e}}{\partial z_{m,e}} (0) = 0,
\end{equation}
\end{enumerate}
Through substitution of equation~(\ref{eq:schrodingersolution}) into equations~(\ref{eq:gaugetransformedcontinuity}) and~(\ref{eq:gaugetransformedkirchoff}), 
we construct the vertex scattering matrix that performs the mapping, 
\begin{equation}
\begin{pmatrix}
\mathbf{a}_\mathcal{L}^{\text{out}} \\
\mathbf{a}_\mathcal{B}^{\text{out}}
\end{pmatrix}
=
S_v
\begin{pmatrix}
\mathbf{a}_\mathcal{L}^{\text{in}} \\
\mathbf{a}_\mathcal{B}^{\text{in}}
\end{pmatrix},
\end{equation}
where,
\begin{equation}
\begin{pmatrix}
\mathbf{a}_\mathcal{L}^{\text{out/in}} \\
\mathbf{a}_\mathcal{B}^{\text{out/in}}
\end{pmatrix} = 
\begin{pmatrix}
a_l^{\text{out/in}} \\
a_r^{\text{out/in}} \\\hline
a_d^{\text{out/in}} \\
a_u^{\text{out/in}} \\
\end{pmatrix}.
\end{equation}
The vertex scattering matrix has elements,
\begin{equation}
S_v=\frac{1}{k + k_p}
\begin{pmatrix}
\begin{array}{cc|cc}
-k_p & k & k_p & k_p\\
    k& -k_p&k_p &k_p\\[.1cm]
    \hline \rule{0cm}{0.5cm}
    k& k&-k& k_p\\
    k& k & k_p & -k\\
    \end{array}
\end{pmatrix}:=
\begin{pmatrix}
S_{\mathcal{LL}} & S_{\mathcal{LB}} \\
S_{\mathcal{BL}} & S_{\mathcal{BB}}
\end{pmatrix}
\end{equation}
Here, $S_{\mathcal{LL}}$ is the scattering matrix that maps wave amplitudes between leads, $S_{\mathcal{LB}}$ maps wave amplitudes between bonds and leads, $S_{\mathcal{BL}}$ maps wave amplitudes between leads and bonds, and $S_{\mathcal{BB}}$ maps wave amplitudes between bonds.

The gauge transformation results in an additional phase $e^{i\phi_p}$ accumulated by the wave as it travels up or down the full length of the waveguide, as determined from the integral in (\ref{eq: Transform}),
\begin{equation}
    \int_{0}^{\ell_{p}} A_{m,u/d} d z_{m,u/d} = \pm\phi_p.
\end{equation}
Formally, 
\begin{equation}\label{eq: edge phases}
\begin{split}
    e^{i \phi_p} \Phi_{m,u} (\ell_{p}) &= \Phi_{m+p,d} (0), \\
    e^{- i \phi_p} \Phi_{m,d} (\ell_{p}) &= \Phi_{m-p,u} (0).
    \end{split}
\end{equation}
By evaluating the phase dynamics along each bond as in (\ref{eq: edge phases}), we write a phase matrix $P$ that performs the mapping,
\begin{equation}\label{eq: Phase Matrix}
\mathbf{a}^{\text{in}}_{\mathcal{B}} = P(k)\mathbf{a}^{\text{out}}_{\mathcal{B}},
\end{equation}
with
\begin{equation}\label{eq:transfer}
    P
    = e^{i k_p \ell_{p}} \begin{pmatrix}
        0 & e^{i \phi_p} e^{- i k_y p \ell} \\
        e^{- i \phi_p} e^{i k_y p \ell} & 0
    \end{pmatrix}.
\end{equation}

We determine the lead scattering matrix $S_{p}$ and thus the scattering properties of the $p^{\text{th}}$ filter layer. Here $S_{p}$ performs the mapping
\begin{equation}
\mathbf{a}^{\text{out}}_{\mathcal{L}}= S_{p} \mathbf{a}^{\text{in}}_{\mathcal{L}}
 = \begin{pmatrix}
        r_{p} & t_{p} \\
        t_{p} & r_{p}
    \end{pmatrix} \mathbf{a}^{\text{in}}_{\mathcal{L}},
\end{equation}
where
\begin{equation}\label{eq: Filter Scattering Matrix}
S_{p} = S_{\mathcal{LL}} + S_{\mathcal{LB}}\left[\mathbf{I} - PS_{\mathcal{BB}}\right]^{-1}PS_{\mathcal{BL}}.
\end{equation}
See \cite{lawrie2023closed} for more details on the above formulation.
\begin{equation}\label{eq: Transmission Definition}
    t_{p} = 
    \frac{i n_p \sin(k n_p \ell_{p})}
    {\cos ( k_y p \ell - \Phi_p) - \cos(k n_p  \ell_{p})  + i n_p \sin(k n_p  \ell_{p})}.
\end{equation}
The reflection coefficient is trivially,
\begin{equation}\label{eq: reflection from transmission}
    r_{p} = t_{p} - 1.
\end{equation}
With that, we have derived the scattering properties of a single layer of the filter in terms of an edge length $\ell_p$, refractive index $n_p$ and magnetic phase $\phi_p$. We now extend this to the scattering properties of a graph formed of an infinite number of bond connections.

\section{Scattering from a Filter Formed of an Infinite Number of Connections.}\label{sec: Scattering from a filter with an infinite number of internal connections}

In the previous section, we determined the scattering properties of a graph filter that connected vertex $m$ to vertex $m \pm p$, via an edge of length $\ell_p$ with refractive index $n_p$ and magnetic phase $\phi_p$. In this section, we determine the scattering properties of a graph filter where every vertex $m$ is connected to every-other vertex.

Suppose we just consider two layers each with index $p = 1$ and $p = 2$. We couple the right leads of layer $1$ to the left leads of layer $2$ transforming the two leads to one finite bidirectional edge of length $\epsilon$ - see Fig. \ref{fig:Graph Setup} for an array of $5$ layers. To model the wave dynamics between the two layers, we introduce the phase $\Psi = \text{e}^{ik\epsilon}$ that performs the mapping
\begin{align}
a_{1,r}^{\text{in}} = \Psi a_{2,l}^{\text{out}} \\
a_{2,l}^{\text{in}} = \Psi a_{1,r}^{\text{out}}
\end{align}
Our aim is to determine the scattering matrix $S_{1,2}$ of the coupled layers, which perform the mapping,
\begin{equation}
\begin{pmatrix}
a_{1,l}^{\text{out}} \\
a_{2,r}^{\text{out}} \\
\end{pmatrix}
 = 
S_{1,2}
\begin{pmatrix}
a_{1,l}^{\text{in}} \\
a_{2,r}^{\text{in}} \\
\end{pmatrix}
=
\begin{pmatrix}
R_{1,1} & T_{1,2} \\
T_{2,1} & R_{2,2}
\end{pmatrix}
\begin{pmatrix}
\mathbf{a}_{1,l}^{\text{in}} \\
\mathbf{a}_{2,r}^{\text{in}} \\
\end{pmatrix}
\end{equation}
Here $R_{n,n}$ is the reflection coefficient at the $n^{\text{th}}$ layer, while $T_{n,m}$ is the transmission coefficient that maps the wave amplitudes from layer $n$ to layer $m$. We determine the transmission and reflection coefficients via a sum over trajectories via a similar method outlined in \cite{andrade2018unitary, DAB19, DAB20}. One can equivelently construct the solutions via equation (\ref{eq: Filter Scattering Matrix}) as in \cite{lawrie2023closed}. Formally, 
\begin{equation}\label{eq: Transmission across two resonators}
\begin{split}
T_{1,2} = T_{2,1} &= t_2 \Psi t_1 \\ 
&+ t_2 \Psi r_2 \Psi r_1 \Psi t_1 \\&+ t_2 \Psi r_2 \Psi r_1 \Psi r_2 \Psi r_1 \Psi t_1  \\
&+ t_2 \Psi r_2 \Psi r_1 \Psi r_2 \Psi r_1 \Psi r_2 \Psi r_1 \Psi r_2 \Psi r_1 \Psi t_1  \\
&+ ... \\
 &=t_2\Psi \sum_{j = 0}^{\infty}\left[r_2 \Psi r_1 \Psi \right]^{j} t_0 \\
 &= \frac{t_2 \Psi t_1}{1 - r_1 r_2\Psi^2}
\end{split}
\end{equation}
The reflection coefficients can easily be determined via the same method.
Suppose now we take the limit as $\epsilon \rightarrow 0^{+}$, then $\Psi \rightarrow 1$, and we remember that the layer reflection coefficient $r_p$ is trivially related to the transmission coefficient via, (\ref{eq: reflection from transmission}), the double layer filter transmission coefficient becomes, 
\begin{equation}
T_{1,2} = T_{2,1} = \frac{1}{\frac{1}{t_1} + \frac{1}{t_2} - 1}.
\end{equation}
It is trivial to extend this formulation to $3$ layers, via the simple recursion relation,
\begin{equation}
\begin{split}
T_{1,3} &= \frac{1}{\frac{1}{T_{1,2}} + \frac{1}{t_3} - 1} \\
&=\frac{1}{\frac{1}{t_1} + \frac{1}{t_2} + \frac{1}{t_3} - 2}.
\end{split}
\end{equation}
This process can continue for an infinite number of connections where we save on indices defining $T:=T_{0,\infty}$,
\begin{equation}
T(k,k_y) = \frac{1}{\sum_{p = 0}^{\infty}\left(\frac{1}{t_p} - 1\right) + 1}
\end{equation}
or equivalently, 
\begin{equation}\label{Total Transmission}
T(k,k_y) = 
\frac{1}{
1 -i \displaystyle\sum_{p = 0}^{\infty}
\frac{\cos(k_y p \ell - \phi_p) - \cos(k n_p \ell_p)}
{ n_p \sin(k n_p \ell_p)}
}.
\end{equation}

The transmission coefficient of Eq.~\eqref{Total Transmission} naturally possesses the structure of a Fourier series through the oscillatory term $\cos(k_y p \ell - \phi_p)$, where each graph connection labelled by $p$ acts as a distinct Fourier harmonic of the filter response. The graph therefore forms a natural platform for Fourier filtering, with the remaining parameters controlling the amplitude, phase and resonance structure of each mode.

\section{Inverse Design of the Filter Transmission Profile}\label{sec: Fourier Representation}

\begin{figure*}[t!]
    \centering
    \includegraphics[width=1\linewidth]{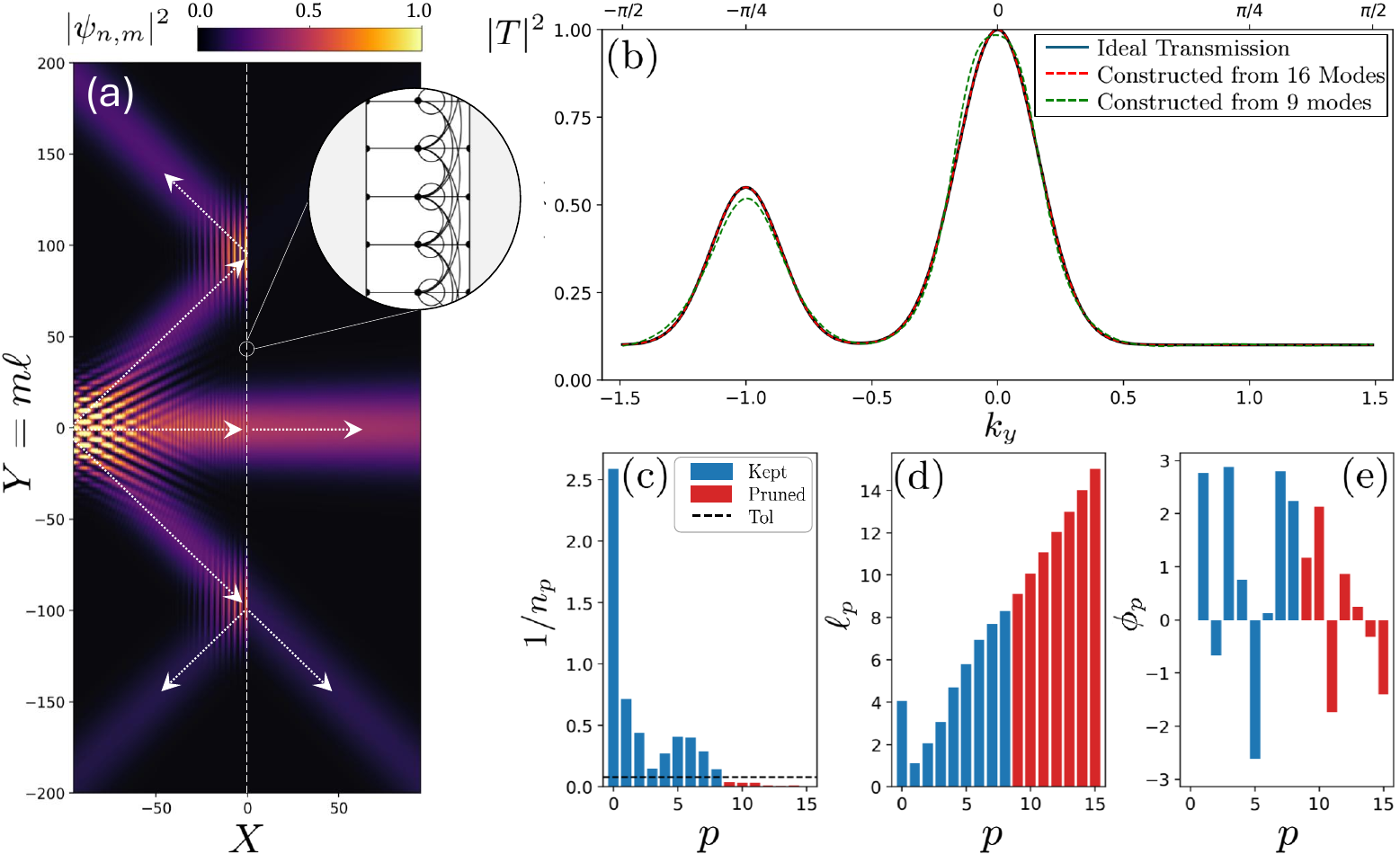}
    \caption{
(a) Norm-squared wave field for three Gaussian beams incident on the graph filter located at $X=0$. The incident beams are centered at angles $\theta = -\pi/4$, $0$ and $\pi/4$, corresponding to the three angular components used to probe the filter response $k_y = -1,0$ and $1$. The inset illustrates the lattice from which the wave solutions are constructed. 
(b) Prescribed transmission intensity and reconstructed graph transmission. The target profile is chosen to give near-unit transmission at normal incidence, reduced transmission at $\pi/4$, and minimal transmission at $-\pi/4$. The full inverse-designed graph reconstruction, the pruned graph reconstruction, and the finite-valency stacked-layer approximation are shown for comparison. 
(c)--(e) Recovered graph parameters for the Fourier components of the filter: inverse bond refractive index $1/n_p$, bond length $\ell_p$, and magnetic phase $\phi_p$, respectively. Blue bars denote retained graph connections, while red bars denote connections removed by the pruning criterion. The pruning demonstrates that weak Fourier components may be discarded, reducing the ideal infinite-valency graph to a finite and physically reasonable set of connections. The finite-valency approximation in (b) corresponds to stacked graph layers, each with reduced valency $d_v = 4+2$, separated by $\epsilon = 0.1\ell$.
}
    \label{fig: 1D example 2}
\end{figure*}

In the previous section we determined the transmission coefficient of a graph filter formed of an infinite number of edge connections, which took the form of the inverse of a series $F$ as in (\ref{Total Transmission}).
Note that this is exactly the form of equation (\ref{Transmission in terms of Reactance}) as derived from the general boundary condition at a thin interface. For passive filters we assume that \(F(k_y;k)\) is real-valued. The transmission intensity can be written trivially as, 
\begin{equation}
|T(k,k_y)|^2
=
\frac{1}{1 + F(k_y;k)^2}.
\end{equation}
This immediately shows that the passive graph construction naturally synthesises a real-valued scattering reactance $F(k_y;k)$ rather than an arbitrary complex transmission coefficient. Consequently, the inverse design problem is most naturally posed in terms of a prescribed transmission intensity profile, where
\begin{equation}
0 < |T(k_y;k)|^2 \leq 1,
\end{equation}

Given a prescribed transmission magnitude profile, the corresponding graph reactance is defined as
\begin{equation}\label{eq:F_from_Tmag}
F(k_y;k)
=
\sqrt{
\frac{1}{|T(k_y;k)|^2} - 1
}.
\end{equation}
By construction, $F(k_y;k)$ is real-valued and therefore admits the Fourier expansion
\begin{equation}\label{eq:F_Fourier}
F(k_y;k)
=
a_0
+
\sum_{p = 1}^{\infty}
\left[
a_p \cos(k_y p \ell)
+
b_p \sin(k_y p \ell)
\right].
\end{equation}
The expansion coefficients are determined via the inverse Fourier transform,
\begin{equation}
\begin{split}
a_0
&=
\frac{\ell}{2\pi}
\int_{-\pi/\ell}^{\pi/\ell}
F(k_y;k)\,dk_y \\
a_p
&=
\frac{\ell}{\pi}
\int_{-\pi/\ell}^{\pi/\ell}
F(k_y;k)\cos(k_y p \ell)\,dk_y, \\
b_p
&=
\frac{\ell}{\pi}
\int_{-\pi/\ell}^{\pi/\ell}
F(k_y;k)\sin(k_y p \ell)\,dk_y.
\end{split}
\end{equation}
In order to construct the above Fourier series from (\ref{Total Transmission}), we impose the edge condition,
\begin{equation}
k n_p \ell_p
=
\pi
\left(
q_p + \frac12
\right),
\quad
q_p \in \mathbb{Z},
\end{equation}
such that
\begin{equation}
\cos(k n_p \ell_p) = 0 \quad \text{and} \quad \sin(k n_p \ell_p)
=
(-1)^{q_p},
\end{equation}
where $q_p$ is some index to be chosen. 
The reactance is then,
\begin{widetext}
\begin{equation}
F(k_y;k)
= \frac{(-1)^{q_0}}{n_0}\cos(\phi_0) + 
\sum_{p = 1}^{\infty}
\frac{(-1)^{q_p}}{n_p}
\left[
\cos(\phi_p)\cos(k_y p \ell)
+
\sin(\phi_p)\sin(k_y p \ell)
\right].
\end{equation}
\end{widetext}
Comparison with equation~(\ref{eq:F_Fourier}) yields, the $p=0$ term
\begin{equation}
a_0 = \frac{(-1)^{q_0}}{n_0}\cos(\phi_0),
\end{equation}
which is independent of $k_y$, supplying the DC offset of the series which physically corresponding to self connected loop at the vertex. For $p\geq 1$,
\begin{equation}\label{expansion}
\begin{split}
a_p
=
\frac{(-1)^{q_p}}{n_p}
\cos(\phi_p),\\
b_p
=
\frac{(-1)^{q_p}}{n_p}
\sin(\phi_p).
\end{split}
\end{equation}
The bond parameters of refractive index, magnetic phase and edge length are then given respectively as,
\begin{equation}
\begin{split}
n_p
&=
\frac{1}{\sqrt{a_p^2+b_p^2}} \\
\phi_p
&=\arctan\left(\frac{b_p}
{a_p}\right) \\
\ell_p
&=
\frac{\pi}{k n_p}
\left(
q_p + \frac12
\right),
\end{split}
\end{equation}
For $p \geq 1$ we physically requiring the bond length to exceed the vertex separation,
\begin{equation}
\ell_p \geq p\ell,
\end{equation}
which yields the constraint on the resonant index,
\begin{equation}
q_p
\geq
\frac{k n_p p \ell}{\pi}
-
\frac12,
\quad
q_p \in \mathbb{Z}.
\end{equation}

Note that $q_p$ can be chosen to ensure that $n_p$ is strictly positive for all $p$ connections.

As an example, consider a transmission profile formed from two Gaussian beams propagating at different angles, each with different amplitudes. The result is plotted in Fig.(\ref{fig: 1D example 2}) for incident beams scattering from the filter. For more information on how to construct beam like solutions, see \cite{lawrie2022quantum} and for information about coupling filters do different media see \cite{lawrie2023engineering} for explicit formulation and \cite{lawrie2025nondiffracting} for a variety of COMSOL simulations. 

Having successfully determined the ideal graph structure from some pre-defined transmission coefficient, we extend this formulation to $2D$ where we determine the ideal graph structure that yields some ideal image.

\section{Extension to a $2D$ Filter}\label{sec: Extension to 2D}

\begin{figure*}[t!]
    \centering
    \includegraphics[width=1\linewidth]{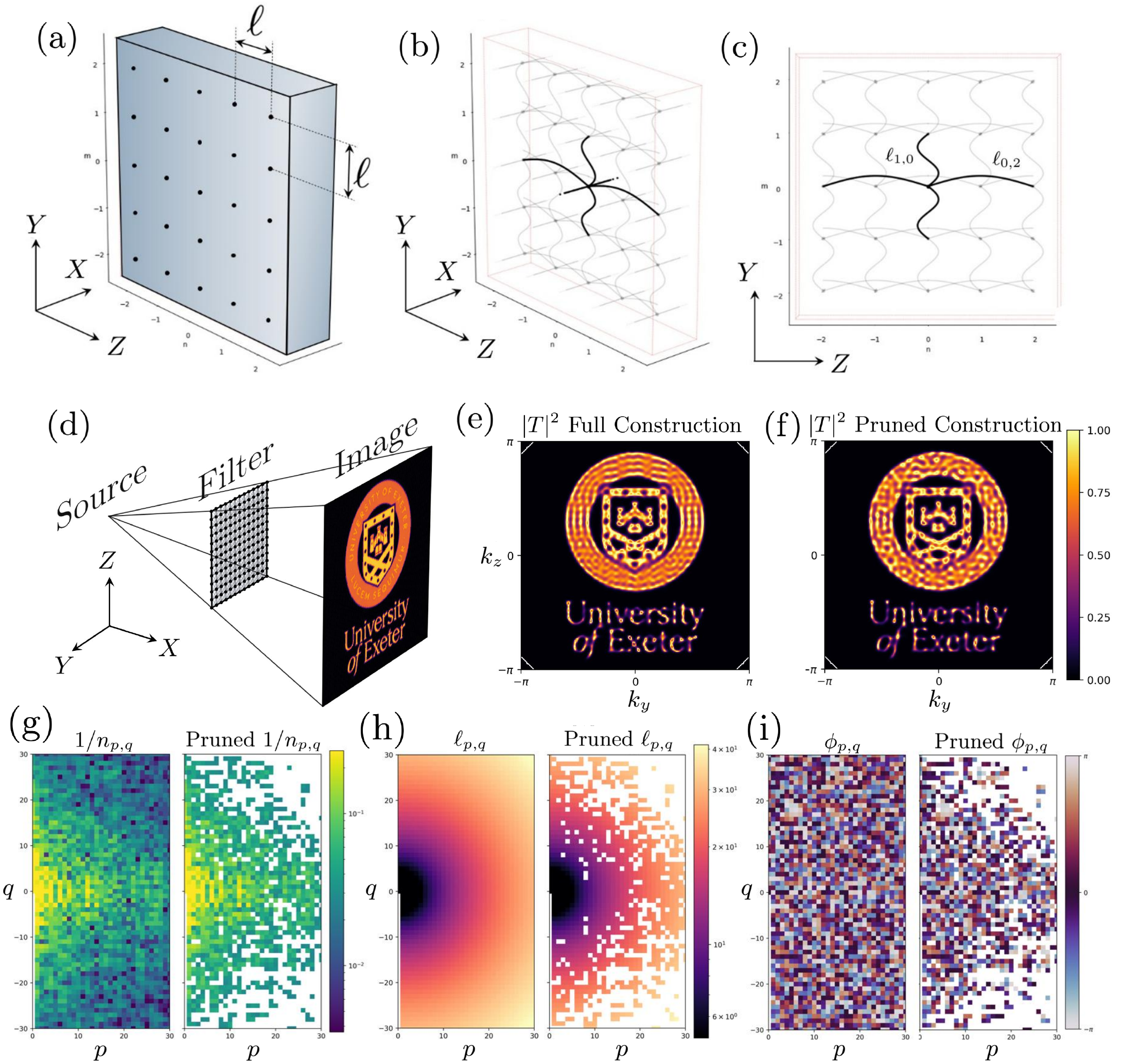}
    \caption{(a) shows a 2D panel with a square array of holes with period $\ell$. (b) shows the inside of the panel formed of thin channels (bonds) that form a periodic graph lattice. Out of the $YZ$ plane are leads that couple the lattice and $3D$ environment. (c) shows a front on view of the panel with edge connections in the horizontal and vertical direction of respective lengths $\ell_{1,0}$ and $\ell_{0,2}$. (d) Schematic of a point-source excitation incident on the two-dimensional filter, producing the University of Exeter logo in the transmitted field. (e) Prescribed transmission intensity $|T(k_y,k_z)|^2$, chosen to encode the University of Exeter logo and reconstructed using $1800$ Fourier modes. (f) Pruned transmission intensity obtained by retaining only modes whose coupling strength lies above a prescribed tolerance. (g)--(i) Recovered graph parameters for the full and pruned filters: inverse bond refractive index $1/n_{p,q}$, bond length $\ell_{p,q}$, and magnetic phase $\phi_{p,q}$, respectively. The comparison shows that weak graph connections may be removed while preserving the dominant features of the prescribed angular transmission profile.}
    \label{fig: 2D Pannel}
\end{figure*}

Extending the above formulation to a $2D$ filter is straightforward. We consider the $2D$ filter as a graph embedded in the $YZ$ plane, with vertices given discrete index $m,n$ spacing with period $\ell$ - see \ref{fig: 2D Pannel} (a - c). The graph bonds now connect vertex $m,n$ to vertex $m \pm p, n \pm q$. For each connection $(p,q)$, we associate a bond length $\ell_{p,q}$, refractive index $n_{p,q}$ and magnetic phase $\phi_{p,q}$. The transmission coefficient is constructed as before via the sum over trajectories, where the graph embedding now spans $\mathbf{R}^2$ and is trivially given as
\begin{widetext}
\begin{equation}
\begin{split}
T(k,k_y,k_z)
&=
\frac{1}{
1-i
\displaystyle
\sum_{(p,q)\in \Omega}
\frac{
\cos(k_y p\ell+k_z q\ell-\phi_{p,q})
-
\cos(k n_{p,q}\ell_{p,q})
}{
 n_{p,q}\sin(k n_{p,q}\ell_{p,q})
}
} \\
\Omega
&=
\{(0,0)\}
\cup
\{(p,q):p>0,\ q\in\mathbb{Z}\}
\cup
\{(0,q):q>0\}.
\end{split}
\end{equation}
\end{widetext}
The discrete index domain $\Omega$ is the half-lattice chosen to avoid double-counting opposite connections. 
As in the one-dimensional case, the transmission coefficient is given in terms of the graph reactance $F$,
\begin{equation}
F(k_y,k_z;k)
=
\sqrt{
\frac{1}{|T(k,k_y,k_z)|^2}-1
},
\end{equation}
which admits the two-dimensional Fourier expansion
\begin{widetext}
\begin{equation}
\begin{split}
F(k_y,k_z;k)
&=
a_{0,0}
+
\sum_{(p,q)\in\mathcal{H}\setminus\{(0,0)\}}
\Big[
a_{p,q}\cos(k_y p\ell+k_z q\ell)
+
b_{p,q}\sin(k_y p\ell+k_z q\ell)
\Big],\\
a_{0,0}
&=
\frac{\ell^2}{4\pi^2}
\int_{-\pi/\ell}^{\pi/\ell}
\int_{-\pi/\ell}^{\pi/\ell}
F(k_y,k_z;k)\,dk_y\,dk_z = \frac{(-1)^{Q_{0,0}}}{n_{0,0}} \text{cos}(\phi_{0,0}), \\
a_{p,q}
&=
\frac{\ell^2}{2\pi^2}
\int_{-\pi/\ell}^{\pi/\ell}
\int_{-\pi/\ell}^{\pi/\ell}
F(k_y,k_z;k)
\cos(k_y p\ell+k_z q\ell)
\,dk_y\,dk_z = \frac{(-1)^{Q_{p,q}}}{n_{p,q}}
\cos(\phi_{p,q}),\\
b_{p,q}
&=
\frac{\ell^2}{2\pi^2}
\int_{-\pi/\ell}^{\pi/\ell}
\int_{-\pi/\ell}^{\pi/\ell}
F(k_y,k_z;k)
\sin(k_y p\ell+k_z q\ell)
\,dk_y\,dk_z =
\frac{(-1)^{Q_{p,q}}}{n_{p,q}}
\sin(\phi_{p,q}).
\end{split}
\end{equation}
\end{widetext}
The bond parameters are then trivially determined in terms of the expansion coefficients giving the bond refractive index, magnetic phase and length as,
\begin{equation}
\begin{split}
n_{p,q}
&=
\frac{1}{\sqrt{a_{p,q}^2+b_{p,q}^2}},\\
\phi_{p,q}
&=
\arctan\left(\frac{b_{p,q}}{a_{p,q}}\right), \\
\ell_{p,q}
&=
\frac{\pi}{k n_{p,q}}
\left(
Q_{p,q}+\frac12
\right).
\end{split}
\end{equation}
For $(p,q)\neq(0,0)$, the physical bond length must exceed the Euclidean separation between the connected vertices,
\begin{equation}
\ell_{p,q}
\geq
\ell\sqrt{p^2+q^2},
\end{equation}
such that the discrete resonant index
\begin{equation}
Q_{p,q}
\geq
\frac{
k n_{p,q}\ell\sqrt{p^2+q^2}
}{\pi}
-
\frac12,
\quad
Q_{p,q}\in\mathbb{Z}^{+}.
\end{equation}

The 2D filter graph parameters are shown in Fig. \ref{fig: 2D Pannel} where we choose a transmission profile of the Exeter University logo.

\section{Conclusion}\label{sec: Conclusion}

In this work we have introduced an analytically invertible framework for wavefront synthesis based upon the scattering properties of periodic waveguide networks. Beginning from the standard interface problem between two semi-infinite half spaces, we showed that the transmission properties of an ideal image-forming interface may be expressed in terms of an effective interface reactance. This formulation establishes a direct relationship between a prescribed transmission coefficient and the boundary conditions required to realise it.

We then introduced a physically realisable implementation of such an interface through a network of thin waveguides, often referred to as a quantum graph. By deriving the exact scattering properties of a graph filter containing an infinite number of beyond-nearest-neighbour connections, we showed that the resulting transmission coefficient naturally takes the form of a Fourier series. Each Fourier mode was found to correspond directly to a graph connection, while the associated Fourier coefficients determine the refractive indices, bond lengths and magnetic phases required to realise the desired response.

This observation allows the inverse problem to be solved analytically. Rather than determining a material architecture through optimisation, machine learning, phase retrieval, topology optimisation or adjoint methods, the desired transmission profile is first specified and then decomposed into its Fourier components. The graph architecture follows directly from these coefficients. In this sense, the network acts as a Fourier synthesiser for transmission coefficients, providing an exact forward construction for image-forming interfaces. The work suggests a shift in perspective for image-forming metamaterials.

\section*{Acknowledgements}
T. M Lawrie would like to thank Simon Horsley, Gregory Chaplain and Gregor Tanner for many useful conversations on this topic. I would also like to thank the financial support of the META4D Program Grant (EP/Y015673/1).

\bibliography{bibliography}% Produces the bibliography via BibTeX.

@article{OvervigYuAlu2021,
  title = {Chiral Quasi-Bound States in the Continuum},
  author = {Overvig, Adam and Yu, Nanfang and Al\`u, Andrea},
  journal = {Phys. Rev. Lett.},
  volume = {126},
  issue = {7},
  pages = {073001},
  numpages = {6},
  year = {2021},
  month = {Feb},
  publisher = {American Physical Society},
  doi = {10.1103/PhysRevLett.126.073001},
  url = {https://link.aps.org/doi/10.1103/PhysRevLett.126.073001}
}

@article{Guo18,
author = {Cheng Guo and Meng Xiao and Momchil Minkov and Yu Shi and Shanhui Fan},
journal = {Optica},
number = {3},
pages = {251--256},
publisher = {Optica Publishing Group},
title = {Photonic crystal slab {Laplace} operator for image differentiation},
volume = {5},
month = {Mar},
year = {2018},
url = {https://opg.optica.org/optica/abstract.cfm?URI=optica-5-3-251},
doi = {10.1364/OPTICA.5.000251},
}

@article{Wang2024,
  title = {Complete inverse design to customize two-dimensional dispersion relation via nonlocal phononic crystals},
  author = {Paul, Sharat and Hasan, Md Nahid and Fu, Henry Chien and Wang, Pai},
  journal = {Phys. Rev. B},
  volume = {110},
  issue = {14},
  pages = {144304},
  numpages = {10},
  year = {2024},
  month = {Oct},
  publisher = {American Physical Society},
  doi = {10.1103/PhysRevB.110.144304},
  url = {https://link.aps.org/doi/10.1103/PhysRevB.110.144304}
}

@article{Silva14,
  title={Performing Mathematical Operations with Metamaterials},
  author={Silva, Alexandre and Monticone, Francesco and Castaldi, Giuseppe and Galdi, Vincenzo and Al\`u, Andrea and Engheta, Nader},
  journal={Science},
  volume={343},
  pages={160--163},
  year={2014}
}

@article{Alu21,
  title={Analogue computing with metamaterials},
  author={Farzad Zangeneth-Nejad and Sounas, Dimitrios L. and Al\`u, Andrea and Fleury Romain},
  journal={Nature Reviews Materials},
  volume={6},
  pages={207--225},
  year={2021}
}

@article{Wesemann21,
  title={Meta-optical and thin film devices for all-optical information processing},
  author={Wesemann, L and Davis, T and Roberts, A},
  journal={Applied Phytsics Reviews},
  volume={8},
  pages={031309},
  year={2021}
}

@article{kottos1999periodic,
  title={Periodic orbit theory and spectral statistics for quantum graphs},
  author={Kottos, Tsampikos and Smilansky, Uzy},
  journal={Annals of Physics},
  volume={274},
  number={1},
  pages={76--124},
  year={1999},
  publisher={Elsevier}
}

@book{berkolaiko2013introduction,
  title={Introduction to quantum graphs},
  author={Berkolaiko, Gregory and Kuchment, Peter},
  number={186},
  year={2013},
  publisher={American Mathematical Soc.}
}

@article{wang2022nonlocal,
  title={Nonlocal interaction engineering of 2D roton-like dispersion relations in acoustic and mechanical metamaterials},
  author={Wang, Ke and Chen, Yi and Kadic, Muamer and Wang, Changguo and Wegener, Martin},
  journal={Communications Materials},
  volume={3},
  number={1},
  pages={35},
  year={2022},
  publisher={Nature Publishing Group UK London}
}

@article{iglesias2021experimental,
  title={Experimental observation of roton-like dispersion relations in metamaterials},
  author={Iglesias Mart{\'\i}nez, Julio Andr{\'e}s and Gro{\ss}, Michael Fidelis and Chen, Yi and Frenzel, Tobias and Laude, Vincent and Kadic, Muamer and Wegener, Martin},
  journal={Science advances},
  volume={7},
  number={49},
  pages={eabm2189},
  year={2021},
  publisher={American Association for the Advancement of Science}
}

@article{chaplain2023reconfigurable,
  title={Reconfigurable elastic metamaterials: Engineering dispersion with beyond nearest neighbors},
  author={Chaplain, GJ and Hooper, IR and Hibbins, AP and Starkey, TA},
  journal={Physical Review Applied},
  volume={19},
  number={4},
  pages={044061},
  year={2023},
  publisher={APS}
}

@article{moore2023acoustic,
  title={Acoustic surface modes on metasurfaces with embedded next-nearest-neighbor coupling},
  author={Moore, DB and Sambles, JR and Hibbins, AP and Starkey, TA and Chaplain, GJ},
  journal={Physical Review B},
  volume={107},
  number={14},
  pages={144110},
  year={2023},
  publisher={APS}
}

@article{chen2023observation,
  title={Observation of Chirality-Induced Roton-Like Dispersion in a 3D Micropolar Elastic Metamaterial},
  author={Chen, Yi and Schneider, Jonathan LG and Gro{\ss}, Michael F and Wang, Ke and Kalt, Sebastian and Scott, Philip and Kadic, Muamer and Wegener, Martin},
  journal={Advanced Functional Materials},
  pages={2302699},
  year={2023},
  publisher={Wiley Online Library}
}

@article{kazemi2023non,
  title={Non-Local Phononic Crystals for Dispersion Customization and Undulation-point Dynamics},
  author={Kazemi, Arash and Deshmukh, Kshiteej J and Chen, Fei and Liu, Yunya and Deng, Bolei and Fu, Henry Chien and Wang, Pai},
  journal={arXiv preprint arXiv:2302.00591},
  year={2023}
}

@article{fleury2021non,
  title={Non-local oddities},
  author={Fleury, Romain},
  journal={Nature Physics},
  volume={17},
  number={7},
  pages={766--767},
  year={2021},
  publisher={Nature Publishing Group UK London}
}

@article{hein2009wave,
  title={Wave communication across regular lattices},
  author={Hein, Birgit and Tanner, Gregor},
  journal={Physical Review Letters},
  volume={103},
  number={26},
  pages={260501},
  year={2009},
  publisher={APS}
}

@article{lawrie2022quantum,
  title={A quantum graph approach to metamaterial design},
  author={Lawrie, Tristan and Tanner, Gregor and Chronopoulos, Dimitrios},
  journal={Scientific Reports},
  volume={12},
  number={1},
  pages={18006},
  year={2022},
  publisher={Nature Publishing Group UK London}
}

@article{kazemi2023drawing,
  title={Drawing Dispersion Curves: Band Structure Customization via Nonlocal Phononic Crystals},
  author={Kazemi, Arash and Deshmukh, Kshiteej J and Chen, Fei and Liu, Yunya and Deng, Bolei and Fu, Henry Chien and Wang, Pai},
  journal={Physical Review Letters},
  volume={131},
  number={17},
  pages={176101},
  year={2023},
  publisher={APS}
}

@article{lawrie2024application,
  title={Application of Quantum Graph Theory to Metamaterial Design: Negative Refraction of Acoustic Waveguide Modes},
  author={Lawrie, TM and Starkey, TA and Tanner, G and Moore, DB and Savage, P and Chaplain, GJ},
  journal={Physical Review Materials (accepted)},
  year={2024}
}

@article{DAB20,
	Author = {Drinko, A. and Andrade, F. M. and Bazeia, D.},
	Da = {2020/06/02},
	Doi = {10.1140/epjp/s13360-020-00459-9},
	Id = {Drinko2020},
	Isbn = {2190-5444},
	Journal = {The European Physical Journal Plus},
	Number = {6},
	Pages = {451},
	Title = {Simple quantum graphs proposal for quantum devices},
	Ty = {JOUR},
	Volume = {135},
	Year = {2020},
		}

@article{kuchment2001convergence,
  title={Convergence of spectra of mesoscopic systems collapsing onto a graph},
  author={Kuchment, Peter and Zeng, Hongbiao},
  journal={Journal of Mathematical Analysis and Applications},
  volume={258},
  number={2},
  pages={671--700},
  year={2001},
  publisher={Elsevier}
}

@phdthesis{lawriethesis,
author = {Lawrie, Tristan},
year = {2025},
month = {07},
pages = {},
title = {PhD Thesis - A Quantum Graph Approach to Metamaterial Design},
doi = {10.13140/RG.2.2.17826.70087}
}

@article{gnutzmann2006quantum,
  title={Quantum graphs: Applications to quantum chaos and universal spectral statistics},
  author={Gnutzmann, Sven and Smilansky, Uzy},
  journal={Advances in Physics},
  volume={55},
  number={5-6},
  pages={527--625},
  year={2006},
  publisher={Taylor \& Francis}
}

@article{edge2025discrete,
  title={Discrete Euler--Bernoulli beam lattices with beyond nearest connections},
  author={Edge, RG and Paul, Erica and Madine, Katie H and Colquitt, Daniel J and Starkey, TA and Chaplain, Gregory James},
  journal={New Journal of Physics},
  volume={27},
  number={2},
  pages={023007},
  year={2025},
  publisher={IOP Publishing}
}

@article{lawrie2023engineering,
  title={Engineering Metamaterial Interface Scattering Coefficients via Quantum Graph Theory.},
  author={Lawrie, Tristan and Tanner, Gregor and Chaplain, Gregory J},
  journal={Acta Physica Polonica: A},
  volume={144},
  number={6},
  year={2023}
}

@article{lawrie2023closed,
  title={Closed form expressions for the {Green’s} function of a quantum graph—a scattering approach},
  author={Lawrie, Tristan and Gnutzmann, Sven and Tanner, Gregor},
  journal={Journal of Physics A: Mathematical and Theoretical},
  volume={56},
  number={47},
  pages={475202},
  year={2023},
  publisher={IOP Publishing}
}

@article{barra2001transport,
  title={Transport and dynamics on open quantum graphs},
  author={Barra, Felipe and Gaspard, Pierre},
  journal={Physical Review E},
  volume={65},
  number={1},
  pages={016205},
  year={2001},
  publisher={APS}
}

@article{DAB19,
  title = {Narrow peaks of full transmission in simple quantum graphs},
  author = {Drinko, A. and Andrade, F. M. and Bazeia, D.},
  journal = {Physical Review A},
  volume = {100},
  issue = {6},
  pages = {062117},
  numpages = {9},
  year = {2019},
  month = {Dec},
  publisher = {American Physical Society}
}

@article{rubinstein2001variational,
  title={Variational Problems on Multiply Connected Thin Strips I: Basic Estimates and Convergence of the Laplacian Spectrum},
  author={Rubinstein, Jacob and Schatzman, Michelle},
  journal={Archive for Rational Mechanics and Analysis},
  volume={160},
  pages={271--308},
  year={2001},
  publisher={Springer}
}

@article{exner2005convergence,
  title={Convergence of spectra of graph-like thin manifolds},
  author={Exner, Pavel and Post, Olaf},
  journal={Journal of Geometry and Physics},
  volume={54},
  number={1},
  pages={77--115},
  year={2005},
  publisher={Elsevier}
}

@article{lawrie2025flux,
  title={A Flux-Tunable Discrete Angular Filter},
  author={Lawrie, TM and Brown, OM},
  journal={Acta Physica Polonica A},
  volume={148},
  number={5},
  pages={S25},
  year={2025}
}

@article{sui2024nonconvex,
  title={Non-convex optimization for inverse problem solving in computer-generated holography},
  author={Sui, Xu and others},
  journal={Light: Science \& Applications},
  volume={13},
  year={2024},
  doi={10.1038/s41377-024-01446-w}
}

@article{yu2025deeplearningcgh,
  title={On the use of deep learning for computer-generated holography},
  author={Yu, Xuan and others},
  journal={iScience},
  volume={28},
  pages={112507},
  year={2025},
  doi={10.1016/j.isci.2025.112507}
}

@article{wang2024phaserecovery,
  title={On the use of deep learning for phase recovery},
  author={Wang, Kai and others},
  journal={Light: Science \& Applications},
  volume={13},
  year={2024},
  doi={10.1038/s41377-023-01340-x}
}

@article{choi2024largearea,
  title={Realization of high-performance optical metasurfaces over a large area: a review from a design perspective},
  author={Choi, M. and others},
  journal={npj Nanophotonics},
  year={2024},
  doi={10.1038/s44310-024-00029-2}
}

@article{zeng2025imagingmetasurface,
  title={From performance to structure: a comprehensive survey of advanced metasurface design for next-generation imaging},
  author={Zeng, Y. and others},
  journal={npj Nanophotonics},
  year={2025},
  doi={10.1038/s44310-025-00081-6}
}

@article{yin2024multidimensional,
  title={Multi-dimensional multiplexed metasurface holography by inverse design},
  author={Yin, Yongyao and others},
  journal={Advanced Materials},
  volume={36},
  pages={2312303},
  year={2024},
  doi={10.1002/adma.202312303}
}

@article{zhang2025multichannel,
  title={End-to-end multichannel holographic metasurface inverse design using an enhanced bi-directional deep neural network},
  author={Zhang, M. and others},
  journal={Optics Express},
  volume={33},
  pages={36801},
  year={2025},
  doi={10.1364/OE.570960}
}

@article{park2025thirtysixchannel,
  title={36-channel spin and wavelength co-multiplexed metasurface holography by phase-gradient inverse design},
  author={Park, Cherry and Jeon, Youngsun and Rho, Junsuk},
  journal={Advanced Science},
  volume={12},
  year={2025},
  doi={10.1002/advs.202504634}
}

@article{yang2025aimetasurface,
  title={Exploring AI in metasurface structures with forward and inverse design},
  author={Yang, Guantai and others},
  journal={iScience},
  volume={28},
  pages={111995},
  year={2025},
  doi={10.1016/j.isci.2025.111995}
}

@article{gabor1948,
  author = {Dennis Gabor},
  title = {A New Microscopic Principle},
  journal = {Nature},
  volume = {161},
  pages = {777--778},
  year = {1948}
}

@article{gerchberg1972,
  author = {R. W. Gerchberg and W. O. Saxton},
  title = {A Practical Algorithm for the Determination of Phase from Image and Diffraction Plane Pictures},
  journal = {Optik},
  volume = {35},
  pages = {237--246},
  year = {1972}
}

@article{yu2011,
  author = {Nanfang Yu and Federico Capasso},
  title = {Flat Optics with Designer Metasurfaces},
  journal = {Nature Materials},
  volume = {13},
  pages = {139--150},
  year = {2014}
}

@article{kildishev2013,
  author = {A. V. Kildishev and A. Boltasseva and V. M. Shalaev},
  title = {Planar Photonics with Metasurfaces},
  journal = {Science},
  volume = {339},
  pages = {1232009},
  year = {2013}
}

@article{padilla2022,
  author = {Willie J. Padilla and Richard D. Averitt},
  title = {Metamaterials for Imaging and Imaging with Metamaterials},
  journal = {Nature Reviews Physics},
  volume = {4},
  pages = {85--100},
  year = {2022}
}

@article{yv76-8rw6,
  title = {Experimental realization of a discrete $k$-space angular filter in microwave networks},
  author = {Ławniczak, Michał and Lawrie, Tristan M. and Bauch, Szymon and Tanner, Gregor and Sirko, Leszek},
  journal = {Phys. Rev. Appl.},
  pages = {},
  year = {2026},
  month = {May},
  publisher = {American Physical Society},
  doi = {10.1103/yv76-8rw6},
  url = {https://link.aps.org/doi/10.1103/yv76-8rw6}
}

@article{andrade2018unitary,
  title={Unitary equivalence between the Green's function and Schr{\"o}dinger approaches for quantum graphs},
  author={Andrade, Fabiano M and Severini, Simone},
  journal={Physical Review A},
  volume={98},
  number={6},
  pages={062107},
  year={2018},
  publisher={APS}
}

@article{edge2025engineering,
  title={Engineering complex dispersion relations},
  author={Edge, RG and Horsley, SAR and Starkey, TA and Chaplain, GJ},
  journal={Physical Review B},
  volume={112},
  number={1},
  pages={014304},
  year={2025},
  publisher={APS}
}

@article{lin2018alloptical,
author = {Lin, Xing and Rivenson, Yair and Yardimci, N. Tolga and Veli, Muhammed and Luo, Yi and Jarrahi, Mona and Ozcan, Aydogan},
title = {All-Optical Machine Learning Using Diffractive Deep Neural Networks},
journal = {Science},
volume = {361},
number = {6406},
pages = {1004--1008},
year = {2018},
doi = {10.1126/science.aat8084}
}

@article{li2022largescale,
author = {Li, Zhaoyi and Pestourie, Rapha{"e}l and Park, Joon-Suh and Huang, Yao-Wei and Johnson, Steven G. and Capasso, Federico},
title = {Inverse Design Enables Large-Scale High-Performance Meta-Optics Reshaping Virtual Reality},
journal = {Nature Communications},
volume = {13},
pages = {2409},
year = {2022},
doi = {10.1038/s41467-022-29973-3}
}

@book{Brillouin1960,
author = {Brillouin, L{'e}on},
title = {Wave Propagation and Group Velocity},
publisher = {Academic Press},
address = {New York},
year = {1960}
}

@article{chen2021roton,
author = {Chen, Yi and Kadic, Muamer and Wegener, Martin},
title = {Roton-Like Acoustical Dispersion Relations in 3D Metamaterials},
journal = {Nature Communications},
volume = {12},
pages = {3278},
year = {2021},
doi = {10.1038/s41467-021-23574-2}
}

@article{lawrie2025nondiffracting,
author = {Lawrie, T. M. and Tanner, G. and Chaplain, G. J.},
title = {Nondiffracting Resonant Angular Filter},
journal = {Physical Review Research},
volume = {7},
pages = {023209},
year = {2025},
doi = {10.1103/PhysRevResearch.7.023209}
}

\end{document}